\documentclass[preprint,aps,showpacs,floatfix]{revtex4}

\usepackage{graphicx}
\usepackage{dcolumn}
\usepackage{bm}
\usepackage{latexsym}

\begin{document}

\title{Strange Quark Contribution to the \\ 
Vector and Axial Form Factors of the Nucleon: \\ 
Combined Analysis of G0, HAPPEx, and Brookhaven E734 Data}

\author{Stephen F. Pate} 
\email{pate@nmsu.edu}
\author{David W. McKee\footnote{Current address: Department of Physics and Astronomy, 
University of Alabama, Tuscaloosa, AL, 35487}}
\author{Vassili Papavassiliou}
\affiliation{Physics Department, New Mexico State University, 
Las Cruces NM 88003}

\date{\today}

\begin{abstract}
The strange quark contribution to the vector and axial form
factors of the nucleon has been determined for momentum transfers in the
range $0.45<Q^2<1.0$ GeV$^2$.  The results are obtained via a combined
analysis of forward-scattering, parity-violating elastic $\vec{e}p$
asymmetry data from the G0 and HAPPEx experiments at Jefferson Lab, 
and elastic
$\nu p$ and $\bar{\nu} p$ scattering data from Experiment 734 at
Brookhaven National Laboratory.  The parity-violating asymmetries
measured in elastic $\vec{e}p$ scattering at forward angles establish
a relationship between the strange vector form factors
$G_E^s$ and $G_M^s$, with little sensitivity to the strange axial form
factor $G_A^s$.  On the other hand, elastic neutrino scattering at low
$Q^2$ is dominated by the axial form factor, with some
significant sensitivity to the vector form factors as well.
Combination of the two data sets allows the simultaneous extraction of
$G_E^s$, $G_M^s$, and $G_A^s$ over a significant range of $Q^2$ for
the very first time.  The $Q^2$-dependence of the strange axial form factor
suggests that the strange quark contribution to the proton spin, 
$\Delta s$, is negative.
\end{abstract}
\maketitle

\section{Historical Motivation}
Ever since the discovery of the first ``strange'' particles in cosmic
ray experiments~\cite{Rochester:1947mi} and the subsequent formulation
of the 3-quark model of baryons~\cite{Gell-Mann:1964nj}, nuclear and
particle physicists have sought to understand the role the strange
quark plays in ``non-strange'' particles like the proton.
Data from deep-inelastic scattering of electrons and
neutrinos from nucleons has determined the unpolarized momentum density
$s(x,Q^2)$ of strange quarks down 
to $x\approx 10^{-4}$~\cite{Pumplin:2002vw, Martin:2001es, Lai:2007dq}. 
Here, $x$ and $Q^2$
are respectively the Bjorken scaling variable and the positive squared
four-momentum transfer for the deep-inelastic scattering vertex.  These
results indicate that the strange quarks account for about 5\% of the
nucleon momentum in the ``infinite momentum frame'' in which these
deep-inelastic data are usually interpreted.  Interest in the role
of the strange quark in the nucleon deepened when the
first polarized inclusive deep-inelastic
measurements of the spin-dependent structure function $g_1(x)$ 
by EMC \cite{Ashman:1987hv,Ashman:1989ig}
demonstrated that the Ellis-Jaffe sum 
rule \cite{Ellis:1973kp,PhysRevD.10.1669.4} did not hold true; these
data implied as well that the strange quark contribution to the proton
spin, $\Delta s$, widely expected to be zero, was possibly negative.  
Subsequent measurements at CERN and SLAC supported the initial EMC 
measurements, and a global analysis \cite{Filippone:2001ux} of these data
suggested $\Delta s \approx -0.15$. 

In the meantime, the E734 experiment \cite{Ahrens:1987xe} at Brookhaven 
measured the $\nu p$ and $\bar{\nu} p$ elastic scattering cross sections
in the momentum-transfer range $0.45<Q^2<1.05$ GeV$^2$.  These cross sections
are very sensitive to the strange axial form factor of the proton,
$G_A^s(Q^2)$, which is related to the strange quark contribution to the
proton spin:  $G_A^s(Q^2=0)=\Delta s$~\cite{Anselmino:1994gn}.
E734 also extracted a negative value for $\Delta s$;  however, this
determination was hampered by the large systematic uncertainties in the
cross section measurement, as well as a lack of knowledge of the strange
vector form factors, and no definitive determination of $\Delta s$ was
possible.  Subsequent reanalyses~\cite{Garvey:1993cg,Alberico:1998qw} confirmed that
the E734 data alone cannot determine $\Delta s$.

More recent measurements using leptonic deep-inelastic scattering have not
given a clear picture of the strange quark contribution to the nucleon spin.
The HERMES experiment~\cite{HERMES_deltas}
measured the helicity distribution
of strange quarks, $\Delta s(x)$, using polarized semi-inclusive deep-inelastic
scattering and a leading-order ``purity'' analysis, and found 
$\Delta s(x)\approx 0$ in the range $0.03<x<0.3$. 
However, a next-to-leading-order analysis
by de Florian, Navarro and Sassot~\cite{deFlorian:2005mw} 
including these HERMES semi-inclusive
data with other world-wide DIS data found a clearly negative strange 
quark polarization in the same $x$-range.  A more recent next-to-leading-order global analysis
including not only leptonic deep-inelastic scattering data but also 
data from the collision of polarized protons at RHIC~\cite{deFlorian:2008mr}
suggests that $\Delta s(x)$ may have a node at $x\approx 0.03$, passing from positive
to negative with decreasing $x$.  The first moment of $\Delta s(x)$ in this fit, giving the 
strange quark spin contribution $\Delta s$, is negative.

In parallel to the effort to determine the activities of the strange quarks
in the proton via deep-inelastic scattering, a strong effort 
has been made to measure the strange quark
contribution to the elastic form factors of the proton, in particular
the vector (electric and magnetic) form factors.  These
experiments~\cite{Mueller:1997mt,Hasty:2001ep,Spayde:2003nr,
Ito:2003mr,Aniol:2004hp,Maas:2004ta,Maas:2004dh,
Armstrong:2005hs,HAPPEx_1H_010,HAPPEx2006} exploit an interference
between the $\gamma$-exchange and $Z$-exchange amplitudes in order to
measure weak elastic form factors $G_E^{Z,p}$ and $G_M^{Z,p}$ which
are the weak-interaction analogs of the more traditional
electromagnetic elastic form factors $G_E^{\gamma,p}$ and
$G_M^{\gamma,p}$ for which copious experimental data are available.
The interference term is observable as a parity-violating asymmetry in
elastic $\vec{e}p$ scattering, with the electron longitudinally
polarized. By combining the electromagnetic form factors of the proton
and neutron with the weak form factors of the proton, one may separate
the up, down, and strange quark contributions.

It is important to point out the differences between what can be measured in
elastic scattering and in different kinds of deep-inelastic scattering, particularly
in regards to the axial form factor.  Elastic scattering of electrons or neutrinos
from nucleons is a neutral current ($\gamma$ or $Z$) process that integrates over
the whole proton wavefunction and so cannot easily 
distinguish between quark and anti-quark.  The strange axial form factor, $G_A^s$,
then is a sum over $s$ and $\bar{s}$ contributions, and likewise the value of
$\Delta s=G_A^s(Q^2=0)$ that might be extracted from those measurements is also
a sum over $s$ and $\bar{s}$.  Similarly, inclusive leptonic deep-inelastic scattering
does not distinguish between quark and anti-quark contributions.  However,
semi-inclusive deep-inelastic scattering, employed by the HERMES
experiment using a model-dependent anaysis involving fragmentation functions, 
can distinguish between quark and anti-quark by observing
leading hardrons in the final state; observation of leading kaons gives HERMES
a window into the contributions of $s$ and $\bar{s}$ quarks to the spin of
the proton.  In this manuscript, we will continue to use the notation $\Delta s$
to refer to the sum of $s$ and $\bar{s}$ contributions to the spin of the
proton, as might be determined from a measurement of $G_A^s$.

With the first HAPPEx~\cite{Aniol:2004hp} measurement of
parity-violating asymmetries in elastic $\vec{e}p$ scattering at
$Q^2=0.477$ GeV$^2$, it became possible to determine
simultaneously~\cite{Pate:2003rk} the strange vector and axial form
factors of the proton by combining those data with the $\nu p$ and
$\bar{\nu} p$ elastic scattering cross sections from Brookhaven E734;
the result was the first determination of the strangeness contribution
to the axial form factor of the proton at non-zero $Q^2$.
Subsequently, additional parity-violating data have become available
from the G0 experiment~\cite{Armstrong:2005hs}.  The purpose of this
paper is to improve and update the analysis done in
Ref.~\cite{Pate:2003rk} by using more complete expressions for the
asymmetries, by including the recent G0 data, by performing a complete
uncertainty analysis, and by comparing the results with available models.

Two global analyses of the parity-violating asymmetries in 
elastic $\vec{e}p$ scattering have been published 
recently~\cite{Young:2006jc,Liu:2007yi}.  The focus of these analyses was
on the vector form factors in the very low $Q^2$ region, $Q^2<0.3$ GeV$^2$, 
with a special interest in extrapolating to $Q^2=0$ to determine 
the strangeness contribution to the proton
magnetic moment.  The emphasis here will be instead on the $Q^2$-dependence
of these form factors, including the strange axial form factor.

\section{Elastic Proton Form Factors: The Strangeness Contribution}
The static properties of the nucleon are described by elastic form factors
defined in terms of matrix elements of current operators.
For example, the matrix element for the electromagnetic
current (one-photon exchange) is expressed as
\begin{eqnarray*}
\raisebox{-.4ex}{${}_N$}\!\left<p'\left| J_\mu^{\gamma}\right|p\right>_N &=&
 \bar{u}(p')\left[\gamma_\mu F_1^{\gamma,N}(Q^2)
  +  i\frac{\sigma_{\mu\nu}q^\nu}{2M}F_2^{\gamma,N}(Q^2)\right]u(p)
\end{eqnarray*}
where the matrix element is taken 
between nucleon states $N$ of momenta $p$ and $p'$,
the momentum transfer is $Q^2=-(p-p')^2$, $u$ is a nucleon spinor,
and $M$ is the mass of the nucleon.  
Similarly, the matrix element of the neutral weak current (one-$Z$ exchange) is
\begin{eqnarray*}
\raisebox{-.4ex}{${}_N$}\!\left<p'\left| J_\mu^{NC}\right|p\right>_N =
\bar{u}(p')\left[\gamma_\mu F_1^{Z,N}(Q^2) \right.
&+& i\frac{\sigma_{\mu\nu}q^\nu}{2M}F_2^{Z,N}(Q^2) \\
&+& \left. \gamma_\mu \gamma_5 G_A^{Z,N}(Q^2)
+\frac{q_\mu}{M}\gamma_5 G_P^{Z,N}(Q^2)\right]u(p).
\end{eqnarray*}
The form factors are respectively the Dirac and Pauli vector
($F_1$ and $F_2$), the axial ($G_A$), and the pseudo-scalar ($G_P$).
Due to the point-like interaction between the gauge bosons ($\gamma$ or $Z$)
and the quarks internal to the nucleon, these form factors can be expressed
as separate contributions from each quark flavor; 
for example, the electromagnetic
and neutral weak Dirac form factors of the proton 
can be expressed in terms of contributions from up, down, and strange quarks:
\begin{eqnarray*}
F_1^{\gamma,p} &=& \frac{2}{3}F_1^u - \frac{1}{3}F_1^d - \frac{1}{3}F_1^s \\
F_1^{Z,p} &=& \left(1-\frac{8}{3}\sin^2\theta_W\right)F_1^u
+\left(-1+\frac{4}{3}\sin^2\theta_W\right)F_1^d
+\left(-1+\frac{4}{3}\sin^2\theta_W\right)F_1^s.
\end{eqnarray*}
The same quark form factors are involved in both expressions; the 
coupling constants that multiply them (electric or weak charges) correspond
to the interaction involved (electromagnetic or weak neutral).
These measurements are most interesting for low momentum 
transfers, $Q^2< 1.0$~GeV$^2$, as the $Q^2=0$ values of these form factors
represent static integral properties of the nucleon.  It is common to use
in these studies the Sachs electric and magnetic form factors
\begin{equation}
G_E = F_1 - \tau F_2 ~~~~~~~~~~ G_M = F_1 + F_2    \nonumber
\end{equation}
instead of the Dirac and Pauli form factors; here, $\tau=Q^2/4M^2$.
At $Q^2=0$ the electromagnetic Sachs electric form factors take on the
value of the nucleon electric charges ($G_E^{\gamma,p}(0)=1$,
$G_E^{\gamma,n}(0)=0$) and the electromagnetic Sachs magnetic form
factors take on the value of the nucleon magnetic moments
($G_M^{\gamma,p}(0)=\mu_p$, $G_M^{\gamma,n}(0)=\mu_n$). Likewise, the
$Q^2=0$ values of the strange quark contributions to these form factors
define the strange contribution to these static quantities: for
example, the strangeness contribution to the proton magnetic moment is
$\mu_s = G_M^s(Q^2=0)$.  It is also common in these studies to assume
charge symmetry; the transformation from proton to neutron form factors 
is an exchange of $u$ and $d$ quark labels.  In addition, it is generally
assumed that the strange quark distributions in the proton and the
neutron are the same.  Then by combining the electromagnetic 
form factors of the proton
and neutron with the weak form factors of the proton, one may separate
the up, down, and strange quark contributions; for example, the
electric form factors may be written as follows:
\begin{eqnarray*}
G_E^{\gamma,p} &=& \frac{2}{3}G_E^u - \frac{1}{3}G_E^d - \frac{1}{3}G_E^s \\
G_E^{\gamma,n} &=& \frac{2}{3}G_E^d - \frac{1}{3}G_E^u - \frac{1}{3}G_E^s \\
G_E^{Z,p} &=& \left(1-\frac{8}{3}\sin^2\theta_W\right)G_E^u
+\left(-1+\frac{4}{3}\sin^2\theta_W\right)G_E^d
+\left(-1+\frac{4}{3}\sin^2\theta_W\right)G_E^s.
\end{eqnarray*}
To attempt this separation is the motivation behind 
the program of parity-violating
$\vec{e}p$ scattering experiments.

The $Z$-exchange current involves also the axial form factor of the
proton, which in a pure weak-interaction process takes
this form:
$$G_A^{Z,p} = \frac{1}{2}\left(-G_A^u + G_A^d + G_A^s\right).$$
The $u-d$ portion of this form factor is well-known from
neutron $\beta$-decay and other charged-current ($CC$) weak 
interaction processes like $\nu_\mu + n \rightarrow p+ \mu^-$:
$$G_A^{CC} = G_A^u-G_A^d = \frac{g_A}{(1+Q^2/M_A^2)^2}$$ 
where $g_A = 1.2695 \pm 0.0029$ is the axial coupling constant in neutron
decay~\cite{PDG2004} and $M_A = 1.001 \pm 0.020$ is the so-called
``axial mass'' which is a fitting parameter for the data on this form
factor~\cite{Budd:2003wb,Bodek:2003ed,Budd:2004bp}.  The strange quark
portion, $G_A^s$, is essentially unknown.  In $\nu p$ and $\bar{\nu}p$
elastic scattering, which are pure neutral-current, weak-interaction
processes, there are no significant radiative corrections to be taken
into account~\cite{Marciano:1980pb}, and we may safely neglect heavy
quark contributions to the axial form factor~\cite{Bass:2002mv}.  On
the other hand, since elastic $ep$ scattering is not a pure
weak-interaction process, then the axial form factor does not appear
in a pure form; there are significant radiative corrections which
carry non-trivial theoretical uncertainties.  The result is that,
while the measurement of parity-violating asymmetries in $\vec{e}p$
elastic scattering is well suited to a measurement of $G_E^s$ and
$G_M^s$, these experiments cannot cleanly extract $G_A^s$.  We will
overcome this difficulty in this analysis by only using
forward-scattering $ep$ data, wherein the axial terms are strongly
suppressed.

\section{Experimental Measurements Sensitive to the Strangeness Form Factors
of the Nucleon} 

There are two principal sources of experimental data from which the
strange quark contribution to the elastic form factors of the proton
may be extracted.  One of these is elastic scattering of neutrinos and
anti-neutrinos from protons; these data are primarily sensitive to the
axial form factor.  The other is the measurement of parity-violating
asymmetries in elastic $\vec{e}p$ scattering; these data are primarily
sensitive to the vector form factors.  This section will describe
these two kinds of experiments.  The following section will describe a
technique to combine these two kinds of data to extract $G_E^s$,
$G_M^s$, and $G_A^s$ simultaneously.

\begin{table}[t]
\caption{\label{table_params} Parameters used in this analysis.
  Uncertainties are listed only if they were of significant size and
  were used to generate the uncertainties in the results. The
  uncertainties on the $R$ factors are not extracted from the
  references, but arise from other considerations; see text for
  details.}
\begin{tabular}{c|c|c}
\hline
Parameter & Value & Reference \\
\hline
$\alpha$ & $7.2973\times 10^{-3}$ & \cite{PDG2004} \\
$\sin^2\theta_W$ & $0.23120$ & \cite{PDG2004} \\
$G_F/(\hbar c)^3$ & $1.16637\times 10^{-5}/{\rm GeV}^2$ & \cite{PDG2004} \\
$g_A$ & $1.2695 \pm 0.0029$ & \cite{PDG2004} \\
$M_A$ & $1.001 \pm 0.020$ GeV & \cite{Budd:2003wb,Bodek:2003ed,Budd:2004bp} \\
$3F-D$ & $0.585\pm 0.025$ & \cite{Goto:1999by} \\
$R^p_V$ & $-0.045 \pm 0.045$ & \cite{PDG2004,Musolf:1994tb}     \\
$R^n_V$ & $-0.012 \pm 0.012 $ & \cite{PDG2004,Musolf:1994tb}     \\
$R^{(0)}_V$ & $-0.012 \pm 0.012 $ & \cite{PDG2004,Musolf:1994tb}     \\
$R^{T=1}_A$ & $-0.173\pm 0.173 $ & \cite{PDG2004,Musolf:1994tb}     \\
$R^{T=0}_A$ & $-0.253\pm 0.253 $ & \cite{PDG2004,Musolf:1994tb}     \\
$R^{(0)}_A$ & $-0.552\pm 0.552 $ & \cite{PDG2004,Musolf:1994tb}     \\
\hline
\end{tabular}
\end{table}

\subsection{Parity-violating Asymmetry in Elastic $\vec{e}p$ Scattering}

The interference between the neutral weak and electromagnetic currents
produces a parity-violating asymmetry in $\vec{e}p$
elastic scattering, which has been the subject of a world-wide
measurement program focussed on the determination of the strange
vector (electric and magnetic) form factors.  
For a proton target, the full expression for the parity-violating
electron scattering asymmetry is \cite{Liu:2007yi,Musolf:1994tb}
\begin{eqnarray}
\label{eqn_asymm}
A^p_{PV} &=& -\frac{G_F Q^2}{4\sqrt{2}\pi\alpha}
               \frac{1}{[\epsilon(G^p_E)^2 + \tau(G^p_M)^2]} \nonumber \\
 &\times& \{(\epsilon(G^p_E)^2+\tau(G^p_M)^2)(1-4\sin^2\theta_W)(1+R^p_V) \nonumber \\
 &      & -(\epsilon G^p_EG^n_E + \tau G^p_MG^n_M)(1+R^n_V) \nonumber \\
 &      & -(\epsilon G^p_EG^s_E + \tau G^p_MG^s_M)(1+R^{(0)}_V) \nonumber \\
 &      & -\epsilon'(1-4\sin^2\theta_W)G^p_MG^e_A\},
\end{eqnarray}
where the kinematics factors are
\begin{eqnarray*}
\epsilon & = & \left[1+2(1+\tau)\tan^2(\theta_e/2)\right]^{-1} \\
\epsilon' & = & \sqrt{(1-\epsilon^2)\tau(1+\tau)}.
\end{eqnarray*}
The axial form factor seen in electron scattering, $G^e_A$, as mentioned earlier,
does not appear in its pure form, but is complicated by radiative
corrections:
\begin{equation}
\label{eqn_ga}
G^e_A(Q^2) = G^{CC}_A(Q^2)(1+R^{T=1}_A)+\sqrt{3}G^8_A(Q^2)R^{T=0}_A+G^s_A(Q^2)(1+R^{(0)}_A).
\end{equation}
The $R$ factors appearing in Equations~\ref{eqn_asymm} and
\ref{eqn_ga} are radiative corrections that may be
expressed~\cite{Musolf:1994tb} in terms of standard model
parameters~\cite{PDG2004}.  Because these radiative corrections are
calculated at $Q^2=0$ and have an unknown $Q^2$-dependence, then in
our analysis some additional uncertainty needs to be attributed to
these radiative correction factors; we have assigned a 100\%
uncertainty to take the unknown $Q^2$-dependence into account (see
Table~\ref{table_params}).  Recently, a reevaluation of these
radiative corrections and their uncertainties, in the context of a fit
to world data on parity-violating $\vec{e}p$ scattering, was discussed
in Ref.~\cite{Liu:2007yi}.  Those values differ from the ones we have
used here; however, the use of these slightly different values would
not have significantly changed the results of the work presented here
because of the supression of the axial terms in the parity-violating
asymmetries at forward angles.

For the vector form factors $G_E^p$, $G_E^n$, $G_M^p$, and $G_M^n$ we
have used the values given by the parametrization of
Kelly~\cite{Kelly:2004hm}.  The uncertainties in the vector form
factors do not contribute significantly to the uncertainties in the
results reported here.  For the charged-current (isovector) axial form
factor, $G_A^{CC}$, as already mentioned, we use a dipole form factor
shape where the $Q^2=0$ value is $g_A = 1.2695 \pm
0.0029$~\cite{PDG2004} and the $Q^2$-dependence is given by the
``axial mass'' parameter $M_A = 1.001 \pm
0.020$~\cite{Budd:2003wb,Bodek:2003ed,Budd:2004bp}.  The selection of
a correct parametrization of $G_A^{CC}$ is crucial to the correct
extraction of $G_A^s$ from neutrino neutral-current data because those
data are sensitive to the total neutral-current axial form factor
$G_A^{Z,p} = (-G_A^{CC} + G_A^s)/2$.  Any shift in the value of
$G_A^{CC}$ will produce a shift in the extracted value of $G_A^s$.  We
chose to use the $M_A$ from
Refs.~\cite{Budd:2003wb,Bodek:2003ed,Budd:2004bp} because they used
up-to-date data on the vector form factors and the value of $g_A$ and
performed a thorough re-evaluation of the original deuterium data on
which the value of $M_A$ is traditionally based.  Recently, two modern
neutrino experiments using nuclear targets (oxygen~\cite{Gran:2006jn}
and carbon~\cite{MB:2007ru}) have reported higher effective values of
$M_A$ from an analysis of charge-current, quasi-elastic scattering.
It not clear at this time what impact these new results have for the
value of $M_A$ for the proton.  If a significantly new set of values
for $G_A^{CC}$ for the proton can be established, then the results for
$G_A^s$ presented in this article will need to be re-evaluated.
In this context it is interesting to note that Kuzmin, Lyubushkin, and
Naumov~\cite{Kuzmin:2007kr} have analyzed a broad range of neutrino
charged-current reaction data, on a wide variety of nuclear targets,
and determined a value for $M_A$ in agreement with
Refs.~\cite{Budd:2003wb,Bodek:2003ed,Budd:2004bp}; this supports our use
of the value $M_A = 1.001 \pm 0.020$. 

Appearing in Equation~\ref{eqn_ga} for $G_A^e$ 
is the octet axial form factor
$G_A^8(Q^2)$.  The $Q^2=0$ value of this form factor is the quantity
$(3F-D)/2\sqrt{3}$; we have taken the value of $3F-D$ from a recent fit
in Ref.~\cite{Goto:1999by} (see Table~\ref{table_params}).  We took the 
$Q^2$-dependence of $G_A^8$ to be the same as that of $G_A^{CC}$, i.e.
$$G_A^8(Q^2) = \frac{(3F-D)/2\sqrt{3}}{(1+Q^2/M_A^2)^2}$$
but this is an assumption.  This form factor is multiplied by the
radiative correction factor $R_A^{T=0}$ to which we have already assigned a
100\% uncertainty because we did not know its $Q^2$-dependence;
as a result, we assigned no additional uncertainty to $G_A^8$.

The parity-violating asymmetry may be written as a linear combination
of the strange electric form factor ($G_E^s$), the strange magnetic
form factor ($G_M^s$), and the strange axial form factor ($G_A^s$), as
follows:
$$A^p_{PV} = A_0^p + A_E^p G_E^s + A_M^p G_M^s + A_A^p G_A^s$$
where the coefficients are
\begin{eqnarray*}
A_0^p & = & -K^p\left\{
\begin{array}{l}
~~~\epsilon G_E^p \left[(1-4\sin^2\theta_W)(1+R_V^p)G_E^p - (1+R_V^n)G_E^n\right] \\
+\tau G_M^p \left[(1-4\sin^2\theta_W)(1+R_V^p)G_M^p - (1+R_V^n)G_M^n\right] \\
-\epsilon' G_M^p(1-4\sin^2\theta_W)
                \left[(1+R_A^{T=1})G^{CC}_A + 
                      \sqrt{3}R_A^{T=0}G_A^8\right] \end{array} \right\} \\
A_E^p & = & K^p\left\{\epsilon G_E^p(1+R_V^0)\right\} \\
A_M^p & = & K^p\left\{\tau G_M^p(1+R_V^0)\right\} \\
A_A^p & = & K^p\left\{\epsilon'G_M^p(1-4\sin^2\theta_W)(1+R_A^0)\right\} \\
K^p & = & \frac{G_F Q^2}{4\pi\sqrt{2}\alpha} \frac{1}{\epsilon {G_E^p}^2 + \tau{G_M^p}^2}.
\end{eqnarray*}
It is well to note that the axial term in this asymmetry
is suppressed by the weak electron charge $(1-4\sin^2\theta_W \approx 0.075)$, and
at forward angles it is suppressed additionally by the kinematic
factor $\epsilon'$.  This might seem a disadvantage, since this
strongly suppresses the sensitivity to the strange axial form factor
in $G_A^e$; however, it simultaneously suppresses the uncertainty in
the radiative corrections in $G_A^e$ which are significant in
magnitude and have an unknown $Q^2$-dependence.  Therefore, the
parity-violating asymmetry data serve to provide a necessary contraint
among the strange vector form factors, with only a little sensitivity
to the strange axial form factor.

\subsection{Cross Section for Elastic $\nu p$ and $\bar{\nu}p$ Scattering}

The neutral weak interaction ($Z$-exchange) process is uniquely
sensitive to the strange axial form factor.  The cross section for
$\nu p$ and $\bar{\nu} p$ elastic scattering is given by
\begin{equation}
\label{eqn_nuxs}
\frac{d\sigma}{dQ^2} = \frac{G_F^2}{2\pi} \frac{Q^2}{E_\nu^2} (A\pm BW + CW^2)
\end{equation}
where the $+$ ($-$) sign is for $\nu$ ($\bar{\nu}$) scattering, and
\begin{eqnarray*}
W    &=& 4(E_\nu /M_p - \tau) \\
\tau &=& Q^2/4M_p^2 \\
A    &=& \frac{1}{4}\left[(G_A^Z)^2(1+\tau)-\left((F_1^Z)^2-\tau(F_2^Z)^2\right)(1-\tau)
          +4\tau{F_1^Z}{F_2^Z}\right] \\
B    &=& -\frac{1}{4}G_A^Z(F_1^Z + F_2^Z) \\
C    &=& \frac{1}{64\tau}\left[(G_A^Z)^2 + (F_1^Z)^2 + \tau(F_2^Z)^2\right].
\end{eqnarray*}
The dependence on the strange form factors is contained in the neutral
current form factors $F_1^Z$, $F_2^Z$, and $G_A^Z$.  At low $Q^2$,
this cross section is dominated by the axial form factor,
\begin{equation}
\frac{d\sigma}{dQ^2}^{\nu p \rightarrow \nu p}(Q^2\rightarrow 0) 
= \frac{G_F^2}{32\pi} \frac{M_p^2}{E_\nu^2}\left[\left(G_A^Z\right)^2 
+ \left(1-4\sin^2\theta_W\right)^2\right], \nonumber
\end{equation}
so these data are a primary source of information about the strange axial
form factor.  Data on these cross sections exists in the range
$0.45<Q^2<1.05$~GeV$^2$ from the Brookhaven E734 
experiment --- see Table~\ref{E734_table}.
Due to a variety of experimental and analytical
difficulties, these data have large total uncertainties, typically 20-25\%.

\begin{table}[t]
\caption{\label{E734_table} Differential cross section data from BNL
  E734~\protect\cite{Ahrens:1987xe}.  The uncertainties shown are
  total; they include statistical, $Q^2$-dependent systematic, and
  $Q^2$-independent systematic contributions, all added in quadrature.
  Also listed is the correlation 
  coefficient $\rho$ for the $\nu$ and $\bar{\nu}$
  data at each value of $Q^2$.}
\begin{tabular}{c|c|c|c}
$Q^2$ & $d\sigma/dQ^2(\nu p)$ & $d\sigma/dQ^2(\bar{\nu} p)$ & Correlation\\
GeV$^2$ &  $10^{-12}$~(fm/GeV)$^2$ & $10^{-12}$~(fm/GeV)$^2$ & Coefficient\\
\hline
0.45 & $0.165  \pm0.033$  & $0.0756 \pm0.0164$ & 0.13 \\
0.55 & $0.109  \pm0.017$  & $0.0426 \pm0.0062$ & 0.26 \\
0.65 & $0.0803 \pm0.0120$ & $0.0283 \pm0.0037$ & 0.29 \\
0.75 & $0.0657 \pm0.0098$ & $0.0184 \pm0.0027$ & 0.26 \\
0.85 & $0.0447 \pm0.0090$ & $0.0129 \pm0.0023$ & 0.16 \\
0.95 & $0.0294 \pm0.0073$ & $0.0108 \pm0.0022$ & 0.12 \\
1.05 & $0.0205 \pm0.0063$ & $0.0101 \pm0.0027$ & 0.07 \\
\end{tabular}
\end{table}

\section{Combined Analysis of BNL E734, HAPPEx, and G0 Data}

A technique for combining forward-scattering parity-violating
$\vec{e}p$ asymmetry data with elastic $\nu p$ and $\bar{\nu} p$ data
was previously used to combine the data from
HAPPEx~\cite{Aniol:2000at,Aniol:2004hp} and Brookhaven Experiment
E734~\cite{Ahrens:1987xe} to extract $G_E^s$, $G_M^s$, $G_A^s$ at
$Q^2=0.5$~GeV$^2 ~ $\cite{Pate:2003rk}.  In the present article we extend
that analysis to include also the recent 
forward-scattering G0 data~\cite{Armstrong:2005hs}
to extract these form factors in the range $0.45<Q^2<1.0$~GeV$^2$.
(Table~\ref{PV_data} lists the
parity-violating asymmetries used in the analysis described here.)
Some small improvements were made in the numerical procedures in the
meantime --- especially, the full expression for the parity-violating asymmetry
$A_{PV}$ is now used, instead of just the linear combination of
$G_E^s$ and $G_M^s$ reported by the experiments as was done in
Ref.~\cite{Pate:2003rk}.  A number of the parameters needed for this
more complete analysis have already been discussed; a list of the parameters
is given in Table~\ref{table_params}.

\begin{table}[t]
\caption{\label{PV_data} Parity-violating asymmetries from the
HAPPEx\cite{Aniol:2000at,Aniol:2004hp} and G0\cite{Armstrong:2005hs} 
experiments that have been used in this analysis.}
\begin{tabular}{c|c|c}
\hline
Experiment & $Q^2$   & $A_{PV}$ \\
           & GeV$^2$ &  ppm \\
\hline
HAPPEx & 0.477 & $-14.92 \pm  1.13 $  \\
    G0 & 0.511 & $-16.81 \pm  2.29 $  \\ 
    G0 & 0.631 & $-19.96 \pm  2.14 $  \\
    G0 & 0.788 & $-30.8  \pm  4.13 $  \\
    G0 & 0.997 & $-37.9  \pm 11.54 $  \\
\hline
\end{tabular}
\end{table}

The data from E734 and that from HAPPEx and G0 are not reported at the
same $Q^2$ values.  The values of $Q^2$ in HAPPEx and G0 are rather
precisely determined, while the results from E734 are averaged over
bins in $Q^2$ that are 0.1 GeV$^2$ wide.  As a result we interpolated
the E734 data (cross-sections, uncertainties, and correlation
coefficients) to the $Q^2$ values of the data from HAPPEx and G0.
Because the E734 results already contain a systematic uncertainty due
to the $Q^2$-determination~\cite{Ahrens:1987xe}, we did not attribute
any additional uncertainty to our interpolation.

Because the neutrino-proton elastic scattering cross sections
(Equation~\ref{eqn_nuxs}) contain quadratic combinations of the
strange form factors, there will be two solutions when these data are
combined wtih the parity-violating asymmetries (which are linear in
the strange form factors, Equation~\ref{eqn_asymm}).  The quadratic
nature of the neutrino data also make difficult an algebraic method of
solution, and so a numerical procedure was developed.

In Table~\ref{solutions_table} the results of the combining of these
data at each value of $Q^2$ are shown.  As mentioned, there are two
solutions at each $Q^2$.  Other available data must be used to
determine which one is the physical sollution.  The two solutions have
distinct features, and there are three strong reasons to prefer
Solution 1:
\begin{itemize}
\item[(1)] The values of $G_E^s$ in Solution 2 are always positive,
  large in magnitude ($\approx 0.3$) and several standard deviations
  away from zero.  However, we do not expect $G_E^s$ to be large; we
  expect it to be small, perhaps zero, since the net electric charge
  from strangeness in the nucleon is zero.  This expectation has been
  borne out in the results of the HAPPEx measurements at $Q^2=0.1$
  GeV$^2$~\cite{HAPPEx2006} which report $G_E^s = -0.005\pm 0.019$.
  In Solution 1, $G_E^s$ is indeed consistent with zero.
\item[(2)] The values of $G_M^s$ in Solution 2 are always negative,
  moderate to large in magnitude ($\approx -0.2$ to $-0.9$), and
  several standard deviations away from zero.  All of the indications
  from experiment so far concerning $G_M^s$ are that it may be small
  and positive, or perhaps zero~\cite{Spayde:2003nr,HAPPEx2006};
  Solution 1 is consistent with these existing indications.
\item[(3)] The values of $G_A^s$ in Solution 2 are always positive,
  moderate in magnitude ($\approx 0.2$), and several standard
  deviations away from zero.  The estimates we have from DIS
  experiments are that $\Delta s = G_A^s(Q^2=0)$ is either zero or
  small and
  negative~\cite{HERMES_deltas,deFlorian:2005mw,deFlorian:2008mr};
  Solution 1 is consistent with those estimates.
\end{itemize}
For these reasons, we select Solution 1 as the physical solution.

\begin{table}[ht]
\caption{\label{solutions_table} Strange form factors for
  $0.45<Q^2<1.0$~GeV$^2$ produced from the E734 and G0 data.  Both
  Solutions 1 and 2 are shown.  Solution 2 is ruled out by other
  experimental data, as explained in the text.}
\begin{tabular}{c||c|c|c||c|c|c}
\hline\hline
  $Q^2$    & \multicolumn{3}{c||}{Solution 1} & \multicolumn{3}{c}{Solution 2} \\
 (GeV$^2$) & $G_E^s$ & $G_M^s$ & $G_A^s$ & $G_E^s$ & $G_M^s$ & $G_A^s$ \\
\hline
0.477 & $~~0.02\pm 0.12$ & $~~0.00\pm 0.29$ & $ -0.127\pm 0.062$ & $~~0.39\pm 0.06$ & $ -0.94\pm 0.15$ & $~~0.266\pm 0.127$ \\ 
0.511 & $~~0.01\pm 0.10$ & $~~0.03\pm 0.22$ & $ -0.103\pm 0.051$ & $~~0.37\pm 0.05$ & $ -0.81\pm 0.12$ & $~~0.257\pm 0.102$ \\
0.631 & $~~0.02\pm 0.07$ & $~~0.08\pm 0.11$ & $ -0.046\pm 0.032$ & $~~0.31\pm 0.03$ & $ -0.47\pm 0.06$ & $~~0.221\pm 0.060$ \\
0.788 & $ -0.02\pm 0.07$ & $~~0.08\pm 0.08$ & $ -0.021\pm 0.029$ & $~~0.25\pm 0.03$ & $ -0.31\pm 0.05$ & $~~0.180\pm 0.050$ \\
0.997 & $ -0.13\pm 0.10$ & $~~0.22\pm 0.07$ & $~~0.015\pm 0.040$ & $~~0.23\pm 0.06$ & $ -0.17\pm 0.06$ & $~~0.225\pm 0.045$ \\
\hline\hline
\end{tabular}
\end{table}

\subsection{Discussion of the Results}

\begin{figure}[t]
\includegraphics[height=.68\textheight, bb=30 190 425 750]{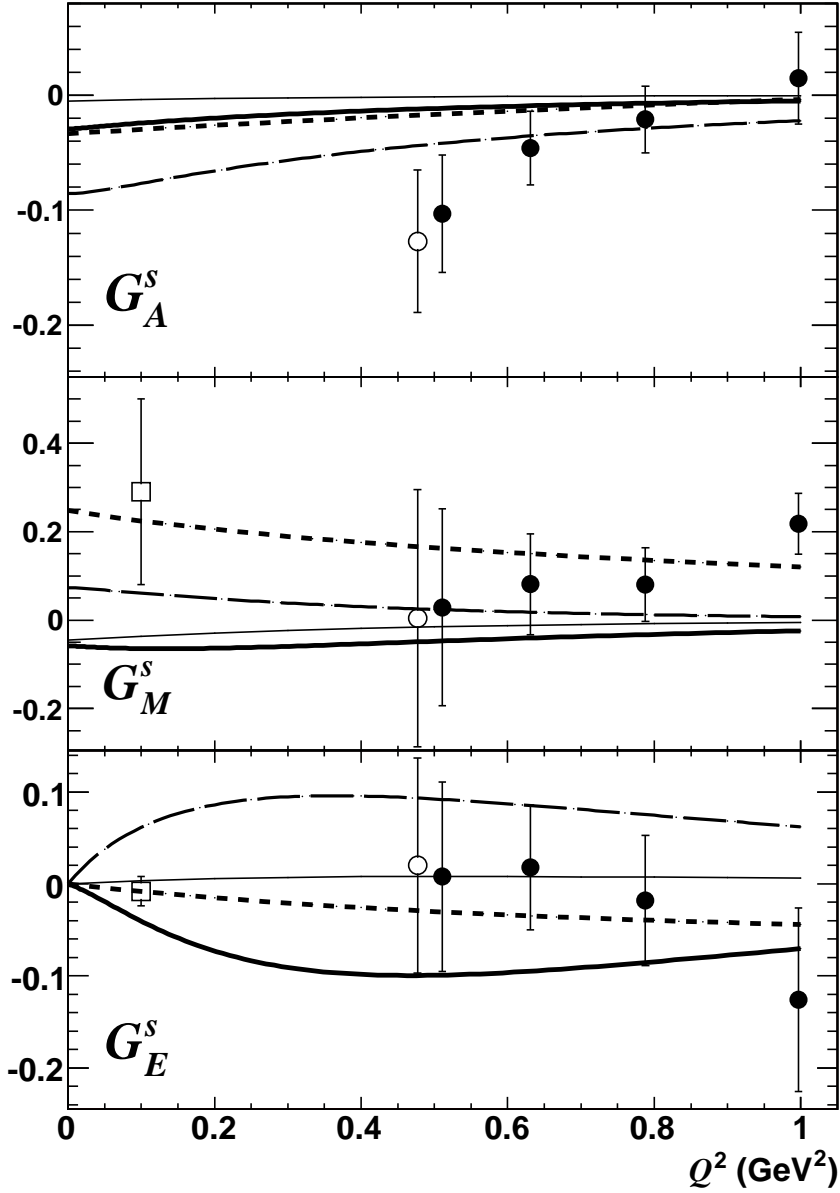}
\caption{Results of this analysis for the strange vector and axial
  form factors of the proton.  Open circles are from a combination of
  HAPPEx and E734 data, while the closed circles are from a
  combination of G0 and E734 data.  [Open squares at $Q^2=0.1$ GeV$^2$
    are from Ref.~\cite{Liu:2007yi} and involve parity-violating
    elastic electron scattering data only.]  The theoretical curves
  are from models that calculate all three of these form factors: Park
  and Weigel~\cite{Park:1991fb} (thick solid line); Lyubovitskij,
  Wang, Gutsche, and Faessler~\cite{Lyubovitskij:2002ng} (thin solid
  line); Silva, Kim, Urbano and
  Goeke~\cite{Silva:2005fa,Goeke:2006gi,Silva:2005qm} (long-dashed
  line); and Riska, An, and
  Zou~\cite{Zou:2005xy,An:2005cj,Riska:2005bh} (short-dashed line).
\label{sff_fig_1}
}
\end{figure}

Figure~\ref{sff_fig_1} displays the results (Solution 1) as a function
of $Q^2$, along with some model calculations that we discuss in
Section~\ref{models}.  Also shown in this figure are the results of
the global analysis by Liu et al.~\cite{Liu:2007yi} of low-$Q^2$
parity-violating electron scattering (PVES) data; that analysis gives
a result for the strange vector form factors $G_E^s$ and $G_M^s$ at
$Q^2$=0.1 GeV$^2$ but not for $G_A^s$ for the reasons discussed
earlier.  (Ref.~\cite{Young:2006jc} has also determined $G_E^s$ and
$G_M^s$ at $Q^2=0.1$ GeV$^2$ using the same data as in
Ref.~\cite{Liu:2007yi}; the results are consistent with those of
Ref.~\cite{Liu:2007yi} but the uncertainties are very much larger.)
The analysis given here, combining a single PVES asymmetry (G0 or
HAPPEx) with the $\nu p$ and $\bar{\nu}p$ data, determines $G_M^s$
with a similar precision as is found from the global fit of multiple
PVES asymmetries.  On the other hand, the PVES data do a spectacular
job in determining $G_E^s$.  The real benefit of the analysis given
here, of course, is the determination of the strangeness contribution
to the axial form factor, $G_A^s$, which shows a trend towards
negative values with decreasing $Q^2$.  This would suggest that
$\Delta s = G_A^2(Q^2=0)$ may be negative as well, but it is clear
from Figure~\ref{sff_fig_1} that the available data do not extend to
sufficiently low $Q^2$ and are not yet precise enough to make a
definitive statement.

\subsection{Correlations among the Results}

It is important to discuss the point-to-point correlations that exist
in these results.  We present these in two different areas:
correlations between different $Q^2$ points for a given form factor,
and correlations between different form factors at a given value of
$Q^2$.

{\bf Correlations for results for a given form factor:} These
correlations arise out of the interpolation of the E734 data points to
the $Q^2$ values of the PV (HAPPEx and G0) data points.  The
interpolation results in a ``sharing'' of E734 points between various
PV data points, and so the results obtained at the PV $Q^2$ points are
not independent of each other.  We express this correlation in terms
of a correlation coefficient, $\rho_{ab} =
\frac{\sigma^2_{ab}}{\sigma_a \sigma_b}$, where $\sigma^2_{ab}$ is the
covariance between results $a$ and $b$, and $\sigma_a$ and $\sigma_b$
are the standard deviations for results $a$ and $b$.  The only
non-zero correlations of this type are between the 0.477 and 0.511
GeV$^2$ points, the 0.477 and 0.631 GeV$^2$ points, and the 0.511 and
0.631 GeV$^2$ points; these are displayed in
Table~\ref{FF_correlation_table}.

\begin{table}[ht]
\caption{\label{FF_correlation_table} Non-zero correlation
  coefficients for different $Q^2$ points for a given form factor.
  The notation $\rho(Q^2_1,Q^2_2)$ refers to a correlation between the
  values of the given form factor at $Q^2_1$ and $Q^2_2$.}
\begin{tabular}{c||c|c|c}
\hline\hline
Form Factor & $\rho(0.477,0.511)$ & $\rho(0.477,0.631)$ & $\rho(0.511,0.631)$ \\
\hline\hline
$G_E^s$ & 0.891 & 0.046 & 0.193 \\
$G_M^s$ & 0.905 & 0.055 & 0.191 \\
$G_A^s$ & 0.884 & 0.038 & 0.189 \\
\hline\hline
\end{tabular}
\end{table}

{\bf Correlations for results for a given $Q^2$:} These correlations
simply express the lack of linear independence of the results, which
is itself due to the algebraic expressions that must be numerically
solved to extract the three strange form factors.  These are displayed
as correlation coefficients in Table~\ref{Q2_correlation_table}; these
correlations are not very strong.

\begin{table}[ht]
\caption{\label{Q2_correlation_table} Correlation coefficients among
the form factors at a given value of $Q^2$.}
\begin{tabular}{c||c|c|c}
\hline\hline
  $Q^2$  & $\rho(G_E^s,G_M^s)$ & $\rho(G_E^s,G_A^s)$ & $\rho(G_M^s,G_A^s)$ \\
 GeV$^2$  &  &  &  \\
\hline\hline
0.477 & -0.199 & 0.060 & -0.061 \\
0.511 & -0.187 & 0.034 & -0.034 \\
0.631 & -0.187 & -0.017 & 0.024 \\
0.788 & -0.169 & -0.049 & 0.064 \\
0.997 & -0.118 & -0.122 & 0.115 \\
\hline\hline
\end{tabular}
\end{table}

\begin{figure}[t]
\includegraphics[height=.68\textheight, bb=30 190 425 750]{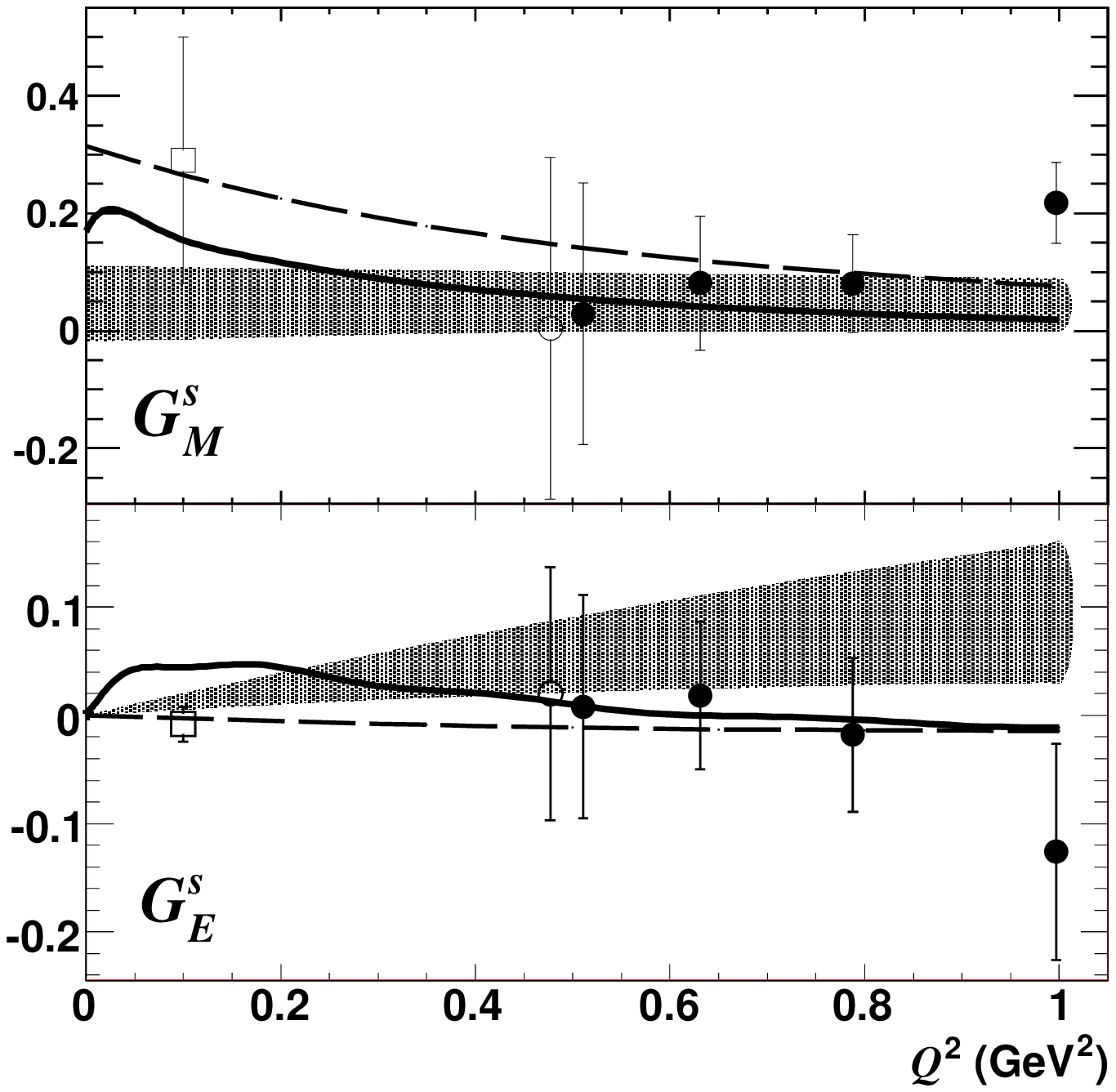}
\caption{Same data as in Figure~\ref{sff_fig_1} for the strange vector
  form factors of the proton, shown in comparison to models which only
  calculate these two form factors: Weigel, Abada, Alkofer and
  Reinhardt~\cite{Weigel:1995jc} (thick solid line);
  Bijker~\cite{Bijker:2005pe} (long-dashed line); and Lewis, Wilcox,
  and Woloshyn~\cite{Lewis:2002ix} (the shaded bands show the upper
  and lower limits of these quantities from this model).
\label{sff_fig_2}
}
\end{figure}

\section{Comparison to Model Calculations}
\label{models}

It is interesting to compare these results with models that can
calculate a $Q^2$-dependence for these form factors.  We distinguish
between those models that provide a calculation of all three form
factors (electric, magnetic and axial --- see Fig.~\ref{sff_fig_1})
and those that only provide a calculation of the vector form factors
(electric and magnetic --- see Fig.~\ref{sff_fig_2}).

Silva, Kim, Urbano and
Goeke~\cite{Silva:2005fa,Goeke:2006gi,Silva:2005qm} have used the
chiral quark soliton model ($\chi$QSM) to calculate $G^s_{E,M,A}(Q^2)$
in the range $0.0<Q^2<1.0$ GeV$^2$.  The $\chi$QSM has been very
successful in reproducing other properties of light baryons using only
a few parameters which are fixed by other data.  In
Figure~\ref{sff_fig_1} their calculation is shown as the long-dashed
line.  Riska, An, and Zou~\cite{Zou:2005xy,An:2005cj,Riska:2005bh}
have explored the strangeness content of the proton by writing all
possible $uuds\bar{s}$ configurations and considering their
contributions to $G^s_{E,M,A}(Q^2)$.  They find that a unique
$uuds\bar{s}$ configuration, with the $s$ quark in a $P$ state and the
$\bar{s}$ in an $S$ state, gives the best fit to the data for these
form factors; see the short-dashed lines in Figure~\ref{sff_fig_1}.
Lyubovitskij, Wang, Gutsche, and Faessler~\cite{Lyubovitskij:2002ng}
have studied these form factors in the framework of the perturbative
chiral quark model; see the thin solid line in Figure~\ref{sff_fig_1}.
Finally, Park and Weigel~\cite{Park:1991fb} have employed the SU(3)
Skyrme model including vector mesons; see the thick solid line in
Figure~\ref{sff_fig_1}.

It is remarkable that none of these models give a satisfactory
description for all three form factors.
Refs.~\cite{Lyubovitskij:2002ng} and
\cite{Zou:2005xy,An:2005cj,Riska:2005bh} give a good description of
the vector form factors, especially the tight constraint placed by the
Liu et al. fit on $G_E^s$ at $Q^2$ = 0.1 GeV$^2$, but fall short of
describing $G_A^s$ well.  On the other hand,
Ref.~\cite{Silva:2005fa,Goeke:2006gi,Silva:2005qm} does the best job
on $G_A^s$ but gives a large value for $G_E^s$ at $Q^2$ = 0.1 GeV$^2$.

Next we consider theoretical models which only provided calculations
of the vector form factors.  Weigel, Abada, Alkofer and
Reinhardt~\cite{Weigel:1995jc} have calculated the strange vector form
factors of the nucleon in the Nambu-Jona-Lasinio soliton model using
the collective approach of Yabu and Ando; their calculation is the
thick solid line in Figure~\ref{sff_fig_2}.
Bijker~\cite{Bijker:2005pe} uses a two-component model of the nucleon
in which the external photon couples to both an intrinsic internal
structure and to a meson cloud through the intermediate vector mesons
($\rho$, $\omega$, $\phi$); the strange quark content comes from the
meson cloud component.  His results are shown as the long-dashed line
in Figure~\ref{sff_fig_2}.  Finally, Lewis, Wilcox, and
Woloshyn~\cite{Lewis:2002ix} have combined quenched lattice QCD
calculations with quenched chiral perturbation theory to calculate the
strangeness contribution to the vector form factors; the shaded bands
in Figure~\ref{sff_fig_2} show the limits on these quantities from
their model.  Two of these models, Refs.~\cite{Bijker:2005pe} and
\cite{Lewis:2002ix}, provide a satisfactory description of the vector
form factors --- it would be interesting to see what these models have
to say about the axial form factor as well.

It is worth noting that Cherman and Cohen~\cite{Cherman:2007qi} have recently
questioned the validity of the calculation of the 
strangeness content of the nucleon from
chiral soliton models similar to those used in 
Refs.~\cite{Silva:2005fa,Goeke:2006gi,Silva:2005qm} and \cite{Weigel:1995jc}.

\section{Future Experiments}

The analysis presented here provides evidence that $G_A^s(Q^2)$
becomes negative with decreasing $Q^2$, implying that $\Delta s =
G_A^s(Q^2=0)$ may itself be negative.  In the near future, the PVA4
experiment~\cite{Baunack:2007zz} will present final backward-angle
PVES data at 0.23 GeV$^2$, and also the G0 experiment will provide
backward-angle PVES data at 0.23 and 0.63 GeV$^2$.  These new data
should greatly constrain the strange vector form factors at these
$Q^2$ points, and will additionally constrain $G_A^s$ to some extent;
the backward-angle data are more sensitive to the axial form factor
than are the forward-angle data.  However, additional (and more
precise) $\nu p$ and $\bar{\nu}p$ elastic scattering data, extended to
lower $Q^2$ values, are needed for a definite determination of the
axial and also the vector form factors.

Several existing and proposed experiments may provide improved
neutrino data in both the near and far term.  The MiniBooNE
experiment~\cite{Cox:2007zz} has already a preliminary result for $\nu
p$ elasic scattering cross sections that could additionally constrain
the strange axial form factor $G_A^s$.  The SciBooNE
experiment~\cite{AlcarazAunion:2007zz}, which has completed a run
already at Fermilab, may be able to provide $\nu p$ elasic scattering
cross sections as well.  FINeSSE~\cite{Bugel:2004yk} proposes to
measure the ratio of the neutral-current to the charged-current $\nu
N$ and $\bar{\nu}N$ processes. A measurement of $R_{NC/CC}=\sigma(\nu
p\rightarrow\nu p)/\sigma(\nu n\rightarrow\mu^- p)$ and
$\bar{R}_{NC/CC}=\sigma(\bar{\nu}p\rightarrow\bar{\nu}p)/\sigma(\bar{\nu}p\rightarrow\mu^+n)$
combined with the world's data on forward-scattering PV $ep$ data can
produce a dense set of data points for $G_A^s$ in the range
$0.25<Q^2<0.75$ GeV$^2$ with an uncertainty at each point of about
$\pm 0.02$.  Another experiment with similar physics goals, called
NeuSpin~\cite{Miyachi:2007zz}, is being proposed for the new JPARC
facility in Japan.

It is also important to extend the semi-inclusive deep-inelastic data
to smaller $x$ and higher $Q^2$ so that the determination of the
polarized strange quark distribution $\Delta s(x)$ can be improved.
The COMPASS experiment~\cite{COMPASSPLANS} will extend the HERMES
measurement of $\Delta s(x)$ in just this way, and a measurement of
this type is also envisioned \cite{Stosslein:2000am,Deshpande:2005wd}
for the proposed electron-ion collider facility.  It is only with
these improved data sets that we will be able to arrive at an
understanding of the strange quark contribution to the proton spin.

The authors are grateful to S.D. Bass, H.E. Jackson, D.O. Riska,
A.W. Thomas, J. Arvieux, E. Beise, E. Leader, R. Young, and
T.W. Donnelly for useful discussions.  This work was supported by the
US Department of Energy, Office of Science.

\bibliography{new_article}

\begin{thebibliography}{63}
\expandafter\ifx\csname natexlab\endcsname\relax\def\natexlab#1{#1}\fi
\expandafter\ifx\csname bibnamefont\endcsname\relax
  \def\bibnamefont#1{#1}\fi
\expandafter\ifx\csname bibfnamefont\endcsname\relax
  \def\bibfnamefont#1{#1}\fi
\expandafter\ifx\csname citenamefont\endcsname\relax
  \def\citenamefont#1{#1}\fi
\expandafter\ifx\csname url\endcsname\relax
  \def\url#1{\texttt{#1}}\fi
\expandafter\ifx\csname urlprefix\endcsname\relax\def\urlprefix{URL }\fi
\providecommand{\bibinfo}[2]{#2}
\providecommand{\eprint}[2][]{\url{#2}}

\bibitem[{\citenamefont{Rochester and Butler}(1947)}]{Rochester:1947mi}
\bibinfo{author}{\bibfnamefont{G.~D.} \bibnamefont{Rochester}}
  \bibnamefont{and} \bibinfo{author}{\bibfnamefont{C.~C.}
  \bibnamefont{Butler}}, \bibinfo{journal}{Nature}
  \textbf{\bibinfo{volume}{160}}, \bibinfo{pages}{855} (\bibinfo{year}{1947}).

\bibitem[{\citenamefont{Gell-Mann}(1964)}]{Gell-Mann:1964nj}
\bibinfo{author}{\bibfnamefont{M.}~\bibnamefont{Gell-Mann}},
  \bibinfo{journal}{Phys. Lett.} \textbf{\bibinfo{volume}{8}},
  \bibinfo{pages}{214} (\bibinfo{year}{1964}).

\bibitem[{\citenamefont{Pumplin et~al.}(2002)}]{Pumplin:2002vw}
\bibinfo{author}{\bibfnamefont{J.}~\bibnamefont{Pumplin}} \bibnamefont{et~al.},
  \bibinfo{journal}{JHEP} \textbf{\bibinfo{volume}{07}}, \bibinfo{pages}{012}
  (\bibinfo{year}{2002}).

\bibitem[{\citenamefont{Martin et~al.}(2002)\citenamefont{Martin, Roberts,
  Stirling, and Thorne}}]{Martin:2001es}
\bibinfo{author}{\bibfnamefont{A.~D.} \bibnamefont{Martin}},
  \bibinfo{author}{\bibfnamefont{R.~G.} \bibnamefont{Roberts}},
  \bibinfo{author}{\bibfnamefont{W.~J.} \bibnamefont{Stirling}},
  \bibnamefont{and} \bibinfo{author}{\bibfnamefont{R.~S.}
  \bibnamefont{Thorne}}, \bibinfo{journal}{Eur. Phys. J.}
  \textbf{\bibinfo{volume}{C23}}, \bibinfo{pages}{73} (\bibinfo{year}{2002}).

\bibitem[{\citenamefont{Lai et~al.}(2007)}]{Lai:2007dq}
\bibinfo{author}{\bibfnamefont{H.~L.} \bibnamefont{Lai}} \bibnamefont{et~al.},
  \bibinfo{journal}{JHEP} \textbf{\bibinfo{volume}{04}}, \bibinfo{pages}{089}
  (\bibinfo{year}{2007}).

\bibitem[{\citenamefont{Ashman et~al.}(1988)}]{Ashman:1987hv}
\bibinfo{author}{\bibfnamefont{J.}~\bibnamefont{Ashman}} \bibnamefont{et~al.}
  (\bibinfo{collaboration}{European Muon Collaboration}),
  \bibinfo{journal}{Phys. Lett.} \textbf{\bibinfo{volume}{B206}},
  \bibinfo{pages}{364} (\bibinfo{year}{1988}).

\bibitem[{\citenamefont{Ashman et~al.}(1989)}]{Ashman:1989ig}
\bibinfo{author}{\bibfnamefont{J.}~\bibnamefont{Ashman}} \bibnamefont{et~al.}
  (\bibinfo{collaboration}{European Muon Collaboration}),
  \bibinfo{journal}{Nucl. Phys.} \textbf{\bibinfo{volume}{B328}},
  \bibinfo{pages}{1} (\bibinfo{year}{1989}).

\bibitem[{\citenamefont{Ellis and Jaffe}(1974{\natexlab{a}})}]{Ellis:1973kp}
\bibinfo{author}{\bibfnamefont{J.~R.} \bibnamefont{Ellis}} \bibnamefont{and}
  \bibinfo{author}{\bibfnamefont{R.~L.} \bibnamefont{Jaffe}},
  \bibinfo{journal}{Phys. Rev.} \textbf{\bibinfo{volume}{D9}},
  \bibinfo{pages}{1444} (\bibinfo{year}{1974}{\natexlab{a}}).

\bibitem[{\citenamefont{Ellis and
  Jaffe}(1974{\natexlab{b}})}]{PhysRevD.10.1669.4}
\bibinfo{author}{\bibfnamefont{J.}~\bibnamefont{Ellis}} \bibnamefont{and}
  \bibinfo{author}{\bibfnamefont{R.}~\bibnamefont{Jaffe}},
  \bibinfo{journal}{Phys. Rev. D} \textbf{\bibinfo{volume}{10}},
  \bibinfo{pages}{1669} (\bibinfo{year}{1974}{\natexlab{b}}).

\bibitem[{\citenamefont{Filippone and Ji}(2001)}]{Filippone:2001ux}
\bibinfo{author}{\bibfnamefont{B.~W.} \bibnamefont{Filippone}}
  \bibnamefont{and} \bibinfo{author}{\bibfnamefont{X.-D.} \bibnamefont{Ji}},
  \bibinfo{journal}{Advances in Nuclear Physics} \textbf{\bibinfo{volume}{26}},
  \bibinfo{pages}{1} (\bibinfo{year}{2001}).

\bibitem[{\citenamefont{Ahrens et~al.}(1987)}]{Ahrens:1987xe}
\bibinfo{author}{\bibfnamefont{L.~A.} \bibnamefont{Ahrens}}
  \bibnamefont{et~al.}, \bibinfo{journal}{Phys. Rev.}
  \textbf{\bibinfo{volume}{D35}}, \bibinfo{pages}{785} (\bibinfo{year}{1987}).

\bibitem[{\citenamefont{Anselmino et~al.}(1995)\citenamefont{Anselmino,
  Efremov, and Leader}}]{Anselmino:1994gn}
\bibinfo{author}{\bibfnamefont{M.}~\bibnamefont{Anselmino}},
  \bibinfo{author}{\bibfnamefont{A.}~\bibnamefont{Efremov}}, \bibnamefont{and}
  \bibinfo{author}{\bibfnamefont{E.}~\bibnamefont{Leader}},
  \bibinfo{journal}{Phys. Rept.} \textbf{\bibinfo{volume}{261}},
  \bibinfo{pages}{1} (\bibinfo{year}{1995}).

\bibitem[{\citenamefont{Garvey et~al.}(1993)\citenamefont{Garvey, Louis, and
  White}}]{Garvey:1993cg}
\bibinfo{author}{\bibfnamefont{G.~T.} \bibnamefont{Garvey}},
  \bibinfo{author}{\bibfnamefont{W.~C.} \bibnamefont{Louis}}, \bibnamefont{and}
  \bibinfo{author}{\bibfnamefont{D.~H.} \bibnamefont{White}},
  \bibinfo{journal}{Phys. Rev.} \textbf{\bibinfo{volume}{C48}},
  \bibinfo{pages}{761} (\bibinfo{year}{1993}).

\bibitem[{\citenamefont{Alberico et~al.}(1999)}]{Alberico:1998qw}
\bibinfo{author}{\bibfnamefont{W.~M.} \bibnamefont{Alberico}}
  \bibnamefont{et~al.}, \bibinfo{journal}{Nucl. Phys.}
  \textbf{\bibinfo{volume}{A651}}, \bibinfo{pages}{277} (\bibinfo{year}{1999}).

\bibitem[{\citenamefont{Airapetian et~al.}(2005)}]{HERMES_deltas}
\bibinfo{author}{\bibfnamefont{A.}~\bibnamefont{Airapetian}}
  \bibnamefont{et~al.} (\bibinfo{collaboration}{HERMES}),
  \bibinfo{journal}{Phys. Rev.} \textbf{\bibinfo{volume}{D71}},
  \bibinfo{pages}{012003} (\bibinfo{year}{2005}).

\bibitem[{\citenamefont{de~Florian et~al.}(2005)\citenamefont{de~Florian,
  Navarro, and Sassot}}]{deFlorian:2005mw}
\bibinfo{author}{\bibfnamefont{D.}~\bibnamefont{de~Florian}},
  \bibinfo{author}{\bibfnamefont{G.~A.} \bibnamefont{Navarro}},
  \bibnamefont{and} \bibinfo{author}{\bibfnamefont{R.}~\bibnamefont{Sassot}},
  \bibinfo{journal}{Phys. Rev.} \textbf{\bibinfo{volume}{D71}},
  \bibinfo{pages}{094018} (\bibinfo{year}{2005}).

\bibitem[{\citenamefont{de~Florian et~al.}(2008)\citenamefont{de~Florian,
  Sassot, Stratmann, and Vogelsang}}]{deFlorian:2008mr}
\bibinfo{author}{\bibfnamefont{D.}~\bibnamefont{de~Florian}},
  \bibinfo{author}{\bibfnamefont{R.}~\bibnamefont{Sassot}},
  \bibinfo{author}{\bibfnamefont{M.}~\bibnamefont{Stratmann}},
  \bibnamefont{and} \bibinfo{author}{\bibfnamefont{W.}~\bibnamefont{Vogelsang}}
  (\bibinfo{year}{2008}), \eprint{arXiv:0804.0422}.

\bibitem[{\citenamefont{Mueller et~al.}(1997)}]{Mueller:1997mt}
\bibinfo{author}{\bibfnamefont{B.}~\bibnamefont{Mueller}} \bibnamefont{et~al.}
  (\bibinfo{collaboration}{SAMPLE}), \bibinfo{journal}{Phys. Rev. Lett.}
  \textbf{\bibinfo{volume}{78}}, \bibinfo{pages}{3824} (\bibinfo{year}{1997}).

\bibitem[{\citenamefont{Hasty et~al.}(2000)}]{Hasty:2001ep}
\bibinfo{author}{\bibfnamefont{R.}~\bibnamefont{Hasty}} \bibnamefont{et~al.}
  (\bibinfo{collaboration}{SAMPLE}), \bibinfo{journal}{Science}
  \textbf{\bibinfo{volume}{290}}, \bibinfo{pages}{2117} (\bibinfo{year}{2000}).

\bibitem[{\citenamefont{Spayde et~al.}(2004)}]{Spayde:2003nr}
\bibinfo{author}{\bibfnamefont{D.~T.} \bibnamefont{Spayde}}
  \bibnamefont{et~al.} (\bibinfo{collaboration}{SAMPLE}),
  \bibinfo{journal}{Phys. Lett.} \textbf{\bibinfo{volume}{B583}},
  \bibinfo{pages}{79} (\bibinfo{year}{2004}).

\bibitem[{\citenamefont{Ito}(2004)}]{Ito:2003mr}
\bibinfo{author}{\bibfnamefont{T.~M.} \bibnamefont{Ito}}
  (\bibinfo{collaboration}{SAMPLE}), \bibinfo{journal}{Phys. Rev. Lett.}
  \textbf{\bibinfo{volume}{92}}, \bibinfo{pages}{102003}
  (\bibinfo{year}{2004}).

\bibitem[{\citenamefont{Aniol et~al.}(2004)}]{Aniol:2004hp}
\bibinfo{author}{\bibfnamefont{K.~A.} \bibnamefont{Aniol}} \bibnamefont{et~al.}
  (\bibinfo{collaboration}{HAPPEX}), \bibinfo{journal}{Phys. Rev.}
  \textbf{\bibinfo{volume}{C69}}, \bibinfo{pages}{065501}
  (\bibinfo{year}{2004}).

\bibitem[{\citenamefont{Maas et~al.}(2004)}]{Maas:2004ta}
\bibinfo{author}{\bibfnamefont{F.~E.} \bibnamefont{Maas}} \bibnamefont{et~al.}
  (\bibinfo{collaboration}{A4}), \bibinfo{journal}{Phys. Rev. Lett.}
  \textbf{\bibinfo{volume}{93}}, \bibinfo{pages}{022002}
  (\bibinfo{year}{2004}).

\bibitem[{\citenamefont{Maas et~al.}(2005)}]{Maas:2004dh}
\bibinfo{author}{\bibfnamefont{F.~E.} \bibnamefont{Maas}} \bibnamefont{et~al.},
  \bibinfo{journal}{Phys. Rev. Lett.} \textbf{\bibinfo{volume}{94}},
  \bibinfo{pages}{152001} (\bibinfo{year}{2005}).

\bibitem[{\citenamefont{Armstrong et~al.}(2005)}]{Armstrong:2005hs}
\bibinfo{author}{\bibfnamefont{D.~S.} \bibnamefont{Armstrong}}
  \bibnamefont{et~al.} (\bibinfo{collaboration}{G0}), \bibinfo{journal}{Phys.
  Rev. Lett.} \textbf{\bibinfo{volume}{95}}, \bibinfo{pages}{092001}
  (\bibinfo{year}{2005}).

\bibitem[{\citenamefont{Aniol et~al.}(2006)}]{HAPPEx_1H_010}
\bibinfo{author}{\bibfnamefont{K.~A.} \bibnamefont{Aniol}} \bibnamefont{et~al.}
  (\bibinfo{collaboration}{HAPPEX}), \bibinfo{journal}{Phys. Lett.}
  \textbf{\bibinfo{volume}{B635}}, \bibinfo{pages}{275} (\bibinfo{year}{2006}).

\bibitem[{\citenamefont{Acha et~al.}(2007)}]{HAPPEx2006}
\bibinfo{author}{\bibfnamefont{A.}~\bibnamefont{Acha}} \bibnamefont{et~al.}
  (\bibinfo{collaboration}{HAPPEX}), \bibinfo{journal}{Phys. Rev. Lett.}
  \textbf{\bibinfo{volume}{98}}, \bibinfo{pages}{032301}
  (\bibinfo{year}{2007}).

\bibitem[{\citenamefont{Pate}(2004)}]{Pate:2003rk}
\bibinfo{author}{\bibfnamefont{S.~F.} \bibnamefont{Pate}},
  \bibinfo{journal}{Phys. Rev. Lett.} \textbf{\bibinfo{volume}{92}},
  \bibinfo{pages}{082002} (\bibinfo{year}{2004}).

\bibitem[{\citenamefont{Young et~al.}(2006)\citenamefont{Young, Roche, Carlini,
  and Thomas}}]{Young:2006jc}
\bibinfo{author}{\bibfnamefont{R.~D.} \bibnamefont{Young}},
  \bibinfo{author}{\bibfnamefont{J.}~\bibnamefont{Roche}},
  \bibinfo{author}{\bibfnamefont{R.~D.} \bibnamefont{Carlini}},
  \bibnamefont{and} \bibinfo{author}{\bibfnamefont{A.~W.}
  \bibnamefont{Thomas}}, \bibinfo{journal}{Phys. Rev. Lett.}
  \textbf{\bibinfo{volume}{97}}, \bibinfo{pages}{102002}
  (\bibinfo{year}{2006}).

\bibitem[{\citenamefont{Liu et~al.}(2007)\citenamefont{Liu, McKeown, and
  Ramsey-Musolf}}]{Liu:2007yi}
\bibinfo{author}{\bibfnamefont{J.}~\bibnamefont{Liu}},
  \bibinfo{author}{\bibfnamefont{R.~D.} \bibnamefont{McKeown}},
  \bibnamefont{and} \bibinfo{author}{\bibfnamefont{M.~J.}
  \bibnamefont{Ramsey-Musolf}}, \bibinfo{journal}{Phys. Rev.}
  \textbf{\bibinfo{volume}{C76}}, \bibinfo{pages}{025202}
  (\bibinfo{year}{2007}).

\bibitem[{\citenamefont{Eidelman et~al.}(2004)}]{PDG2004}
\bibinfo{author}{\bibfnamefont{S.}~\bibnamefont{Eidelman}} \bibnamefont{et~al.}
  (\bibinfo{collaboration}{Particle Data Group}), \bibinfo{journal}{Phys.
  Lett.} \textbf{\bibinfo{volume}{B592}}, \bibinfo{pages}{1}
  (\bibinfo{year}{2004}).

\bibitem[{\citenamefont{Bodek et~al.}(2004)\citenamefont{Bodek, Budd, and
  Arrington}}]{Bodek:2003ed}
\bibinfo{author}{\bibfnamefont{A.}~\bibnamefont{Bodek}},
  \bibinfo{author}{\bibfnamefont{H.}~\bibnamefont{Budd}}, \bibnamefont{and}
  \bibinfo{author}{\bibfnamefont{J.}~\bibnamefont{Arrington}},
  \bibinfo{journal}{AIP Conf. Proc.} \textbf{\bibinfo{volume}{698}},
  \bibinfo{pages}{148} (\bibinfo{year}{2004}).

\bibitem[{\citenamefont{Budd et~al.}(2003)\citenamefont{Budd, Bodek, and
  Arrington}}]{Budd:2003wb}
\bibinfo{author}{\bibfnamefont{H.}~\bibnamefont{Budd}},
  \bibinfo{author}{\bibfnamefont{A.}~\bibnamefont{Bodek}}, \bibnamefont{and}
  \bibinfo{author}{\bibfnamefont{J.}~\bibnamefont{Arrington}}
  (\bibinfo{year}{2003}), \eprint{hep-ex/0308005}.

\bibitem[{\citenamefont{Budd et~al.}(2005)\citenamefont{Budd, Bodek, and
  Arrington}}]{Budd:2004bp}
\bibinfo{author}{\bibfnamefont{H.}~\bibnamefont{Budd}},
  \bibinfo{author}{\bibfnamefont{A.}~\bibnamefont{Bodek}}, \bibnamefont{and}
  \bibinfo{author}{\bibfnamefont{J.}~\bibnamefont{Arrington}},
  \bibinfo{journal}{Nucl. Phys. Proc. Suppl.} \textbf{\bibinfo{volume}{139}},
  \bibinfo{pages}{90} (\bibinfo{year}{2005}).

\bibitem[{\citenamefont{Marciano and Sirlin}(1980)}]{Marciano:1980pb}
\bibinfo{author}{\bibfnamefont{W.~J.} \bibnamefont{Marciano}} \bibnamefont{and}
  \bibinfo{author}{\bibfnamefont{A.}~\bibnamefont{Sirlin}},
  \bibinfo{journal}{Phys. Rev.} \textbf{\bibinfo{volume}{D22}},
  \bibinfo{pages}{2695} (\bibinfo{year}{1980}).

\bibitem[{\citenamefont{Bass et~al.}(2002)\citenamefont{Bass, Crewther,
  Steffens, and Thomas}}]{Bass:2002mv}
\bibinfo{author}{\bibfnamefont{S.~D.} \bibnamefont{Bass}},
  \bibinfo{author}{\bibfnamefont{R.~J.} \bibnamefont{Crewther}},
  \bibinfo{author}{\bibfnamefont{F.~M.} \bibnamefont{Steffens}},
  \bibnamefont{and} \bibinfo{author}{\bibfnamefont{A.~W.}
  \bibnamefont{Thomas}}, \bibinfo{journal}{Phys. Rev.}
  \textbf{\bibinfo{volume}{D66}}, \bibinfo{pages}{031901}
  (\bibinfo{year}{2002}).

\bibitem[{\citenamefont{Goto et~al.}(2000)}]{Goto:1999by}
\bibinfo{author}{\bibfnamefont{Y.}~\bibnamefont{Goto}} \bibnamefont{et~al.}
  (\bibinfo{collaboration}{Asymmetry Analysis}), \bibinfo{journal}{Phys. Rev.}
  \textbf{\bibinfo{volume}{D62}}, \bibinfo{pages}{034017}
  (\bibinfo{year}{2000}).

\bibitem[{\citenamefont{Musolf et~al.}(1994)}]{Musolf:1994tb}
\bibinfo{author}{\bibfnamefont{M.~J.} \bibnamefont{Musolf}}
  \bibnamefont{et~al.}, \bibinfo{journal}{Phys. Rept.}
  \textbf{\bibinfo{volume}{239}}, \bibinfo{pages}{1} (\bibinfo{year}{1994}).

\bibitem[{\citenamefont{Kelly}(2004)}]{Kelly:2004hm}
\bibinfo{author}{\bibfnamefont{J.~J.} \bibnamefont{Kelly}},
  \bibinfo{journal}{Phys. Rev.} \textbf{\bibinfo{volume}{C70}},
  \bibinfo{pages}{068202} (\bibinfo{year}{2004}).

\bibitem[{\citenamefont{Gran et~al.}(2006)}]{Gran:2006jn}
\bibinfo{author}{\bibfnamefont{R.}~\bibnamefont{Gran}} \bibnamefont{et~al.}
  (\bibinfo{collaboration}{K2K}), \bibinfo{journal}{Phys. Rev.}
  \textbf{\bibinfo{volume}{D74}}, \bibinfo{pages}{052002}
  (\bibinfo{year}{2006}).

\bibitem[{\citenamefont{Aguilar-Arevalo et~al.}(2008)}]{MB:2007ru}
\bibinfo{author}{\bibfnamefont{A.~A.} \bibnamefont{Aguilar-Arevalo}}
  \bibnamefont{et~al.} (\bibinfo{collaboration}{MiniBooNE}),
  \bibinfo{journal}{Phys. Rev. Lett.} \textbf{\bibinfo{volume}{100}},
  \bibinfo{pages}{032301} (\bibinfo{year}{2008}).

\bibitem[{\citenamefont{Kuzmin et~al.}(2008)\citenamefont{Kuzmin, Lyubushkin,
  and Naumov}}]{Kuzmin:2007kr}
\bibinfo{author}{\bibfnamefont{K.~S.} \bibnamefont{Kuzmin}},
  \bibinfo{author}{\bibfnamefont{V.~V.} \bibnamefont{Lyubushkin}},
  \bibnamefont{and} \bibinfo{author}{\bibfnamefont{V.~A.}
  \bibnamefont{Naumov}}, \bibinfo{journal}{Eur. Phys. J.}
  \textbf{\bibinfo{volume}{C54}}, \bibinfo{pages}{517} (\bibinfo{year}{2008}).

\bibitem[{\citenamefont{Aniol et~al.}(2001)}]{Aniol:2000at}
\bibinfo{author}{\bibfnamefont{K.~A.} \bibnamefont{Aniol}} \bibnamefont{et~al.}
  (\bibinfo{collaboration}{HAPPEX}), \bibinfo{journal}{Phys. Lett.}
  \textbf{\bibinfo{volume}{B509}}, \bibinfo{pages}{211} (\bibinfo{year}{2001}).

\bibitem[{\citenamefont{Park and Weigel}(1992)}]{Park:1991fb}
\bibinfo{author}{\bibfnamefont{N.~W.} \bibnamefont{Park}} \bibnamefont{and}
  \bibinfo{author}{\bibfnamefont{H.}~\bibnamefont{Weigel}},
  \bibinfo{journal}{Nucl. Phys.} \textbf{\bibinfo{volume}{A541}},
  \bibinfo{pages}{453} (\bibinfo{year}{1992}).

\bibitem[{\citenamefont{Lyubovitskij et~al.}(2002)\citenamefont{Lyubovitskij,
  Wang, Gutsche, and Faessler}}]{Lyubovitskij:2002ng}
\bibinfo{author}{\bibfnamefont{V.~E.} \bibnamefont{Lyubovitskij}},
  \bibinfo{author}{\bibfnamefont{P.}~\bibnamefont{Wang}},
  \bibinfo{author}{\bibfnamefont{T.}~\bibnamefont{Gutsche}}, \bibnamefont{and}
  \bibinfo{author}{\bibfnamefont{A.}~\bibnamefont{Faessler}},
  \bibinfo{journal}{Phys. Rev.} \textbf{\bibinfo{volume}{C66}},
  \bibinfo{pages}{055204} (\bibinfo{year}{2002}).

\bibitem[{\citenamefont{Silva et~al.}(2005)\citenamefont{Silva, Kim, Urbano,
  and Goeke}}]{Silva:2005fa}
\bibinfo{author}{\bibfnamefont{A.}~\bibnamefont{Silva}},
  \bibinfo{author}{\bibfnamefont{H.-C.} \bibnamefont{Kim}},
  \bibinfo{author}{\bibfnamefont{D.}~\bibnamefont{Urbano}}, \bibnamefont{and}
  \bibinfo{author}{\bibfnamefont{K.}~\bibnamefont{Goeke}},
  \bibinfo{journal}{Phys. Rev.} \textbf{\bibinfo{volume}{D72}},
  \bibinfo{pages}{094011} (\bibinfo{year}{2005}).

\bibitem[{\citenamefont{Goeke et~al.}(2007)\citenamefont{Goeke, Kim, Silva, and
  Urbano}}]{Goeke:2006gi}
\bibinfo{author}{\bibfnamefont{K.}~\bibnamefont{Goeke}},
  \bibinfo{author}{\bibfnamefont{H.-C.} \bibnamefont{Kim}},
  \bibinfo{author}{\bibfnamefont{A.}~\bibnamefont{Silva}}, \bibnamefont{and}
  \bibinfo{author}{\bibfnamefont{D.}~\bibnamefont{Urbano}},
  \bibinfo{journal}{Eur. Phys. J.} \textbf{\bibinfo{volume}{A32}},
  \bibinfo{pages}{393} (\bibinfo{year}{2007}).

\bibitem[{\citenamefont{Silva et~al.}(2006)\citenamefont{Silva, Kim, Urbano,
  and Goeke}}]{Silva:2005qm}
\bibinfo{author}{\bibfnamefont{A.}~\bibnamefont{Silva}},
  \bibinfo{author}{\bibfnamefont{H.-C.} \bibnamefont{Kim}},
  \bibinfo{author}{\bibfnamefont{D.}~\bibnamefont{Urbano}}, \bibnamefont{and}
  \bibinfo{author}{\bibfnamefont{K.}~\bibnamefont{Goeke}},
  \bibinfo{journal}{Phys. Rev.} \textbf{\bibinfo{volume}{D74}},
  \bibinfo{pages}{054011} (\bibinfo{year}{2006}).

\bibitem[{\citenamefont{Zou and Riska}(2005)}]{Zou:2005xy}
\bibinfo{author}{\bibfnamefont{B.~S.} \bibnamefont{Zou}} \bibnamefont{and}
  \bibinfo{author}{\bibfnamefont{D.~O.} \bibnamefont{Riska}},
  \bibinfo{journal}{Phys. Rev. Lett.} \textbf{\bibinfo{volume}{95}},
  \bibinfo{pages}{072001} (\bibinfo{year}{2005}).

\bibitem[{\citenamefont{An et~al.}(2006)\citenamefont{An, Riska, and
  Zou}}]{An:2005cj}
\bibinfo{author}{\bibfnamefont{C.~S.} \bibnamefont{An}},
  \bibinfo{author}{\bibfnamefont{D.~O.} \bibnamefont{Riska}}, \bibnamefont{and}
  \bibinfo{author}{\bibfnamefont{B.~S.} \bibnamefont{Zou}},
  \bibinfo{journal}{Phys. Rev.} \textbf{\bibinfo{volume}{C73}},
  \bibinfo{pages}{035207} (\bibinfo{year}{2006}).

\bibitem[{\citenamefont{Riska and Zou}(2006)}]{Riska:2005bh}
\bibinfo{author}{\bibfnamefont{D.~O.} \bibnamefont{Riska}} \bibnamefont{and}
  \bibinfo{author}{\bibfnamefont{B.~S.} \bibnamefont{Zou}},
  \bibinfo{journal}{Phys. Lett.} \textbf{\bibinfo{volume}{B636}},
  \bibinfo{pages}{265} (\bibinfo{year}{2006}).

\bibitem[{\citenamefont{Weigel et~al.}(1995)\citenamefont{Weigel, Abada,
  Alkofer, and Reinhardt}}]{Weigel:1995jc}
\bibinfo{author}{\bibfnamefont{H.}~\bibnamefont{Weigel}},
  \bibinfo{author}{\bibfnamefont{A.}~\bibnamefont{Abada}},
  \bibinfo{author}{\bibfnamefont{R.}~\bibnamefont{Alkofer}}, \bibnamefont{and}
  \bibinfo{author}{\bibfnamefont{H.}~\bibnamefont{Reinhardt}},
  \bibinfo{journal}{Phys. Lett.} \textbf{\bibinfo{volume}{B353}},
  \bibinfo{pages}{20} (\bibinfo{year}{1995}).

\bibitem[{\citenamefont{Bijker}(2006)}]{Bijker:2005pe}
\bibinfo{author}{\bibfnamefont{R.}~\bibnamefont{Bijker}}, \bibinfo{journal}{J.
  Phys.} \textbf{\bibinfo{volume}{G32}}, \bibinfo{pages}{L49}
  (\bibinfo{year}{2006}).

\bibitem[{\citenamefont{Lewis et~al.}(2003)\citenamefont{Lewis, Wilcox, and
  Woloshyn}}]{Lewis:2002ix}
\bibinfo{author}{\bibfnamefont{R.}~\bibnamefont{Lewis}},
  \bibinfo{author}{\bibfnamefont{W.}~\bibnamefont{Wilcox}}, \bibnamefont{and}
  \bibinfo{author}{\bibfnamefont{R.~M.} \bibnamefont{Woloshyn}},
  \bibinfo{journal}{Phys. Rev.} \textbf{\bibinfo{volume}{D67}},
  \bibinfo{pages}{013003} (\bibinfo{year}{2003}).

\bibitem[{\citenamefont{Cherman and Cohen}(2007)}]{Cherman:2007qi}
\bibinfo{author}{\bibfnamefont{A.}~\bibnamefont{Cherman}} \bibnamefont{and}
  \bibinfo{author}{\bibfnamefont{T.~D.} \bibnamefont{Cohen}},
  \bibinfo{journal}{Phys. Lett.} \textbf{\bibinfo{volume}{B651}},
  \bibinfo{pages}{39} (\bibinfo{year}{2007}).

\bibitem[{\citenamefont{Baunack}(2007)}]{Baunack:2007zz}
\bibinfo{author}{\bibfnamefont{S.}~\bibnamefont{Baunack}},
  \bibinfo{journal}{Eur. Phys. J.} \textbf{\bibinfo{volume}{A32}},
  \bibinfo{pages}{457} (\bibinfo{year}{2007}).

\bibitem[{\citenamefont{Cox}(2007)}]{Cox:2007zz}
\bibinfo{author}{\bibfnamefont{D.~C.} \bibnamefont{Cox}}
  (\bibinfo{collaboration}{MiniBooNE}), \bibinfo{journal}{AIP Conf. Proc.}
  \textbf{\bibinfo{volume}{967}}, \bibinfo{pages}{130} (\bibinfo{year}{2007}).

\bibitem[{\citenamefont{Alcaraz-Aunion and
  Catala-Perez}(2007)}]{AlcarazAunion:2007zz}
\bibinfo{author}{\bibfnamefont{J.~L.} \bibnamefont{Alcaraz-Aunion}}
  \bibnamefont{and}
  \bibinfo{author}{\bibfnamefont{J.}~\bibnamefont{Catala-Perez}},
  \bibinfo{journal}{AIP Conf. Proc.} \textbf{\bibinfo{volume}{967}},
  \bibinfo{pages}{307} (\bibinfo{year}{2007}).

\bibitem[{\citenamefont{Bugel et~al.}(2004)}]{Bugel:2004yk}
\bibinfo{author}{\bibfnamefont{L.}~\bibnamefont{Bugel}} \bibnamefont{et~al.}
  (\bibinfo{collaboration}{FINeSSE}) (\bibinfo{year}{2004}),
  \eprint{hep-ex/0402007}.

\bibitem[{\citenamefont{Miyachi et~al.}(2007)\citenamefont{Miyachi, Shibata,
  and Saito}}]{Miyachi:2007zz}
\bibinfo{author}{\bibfnamefont{Y.}~\bibnamefont{Miyachi}},
  \bibinfo{author}{\bibfnamefont{T.-A.} \bibnamefont{Shibata}},
  \bibnamefont{and} \bibinfo{author}{\bibfnamefont{N.}~\bibnamefont{Saito}},
  \bibinfo{journal}{AIP Conf. Proc.} \textbf{\bibinfo{volume}{915}},
  \bibinfo{pages}{387} (\bibinfo{year}{2007}).

\bibitem[{COM(2007)}]{COMPASSPLANS}
\bibinfo{type}{CERN Report} \bibinfo{number}{CERN-SPSC-2007-002}
  (\bibinfo{year}{2007}).

\bibitem[{\citenamefont{St{\"o}sslein and Kinney}(2001)}]{Stosslein:2000am}
\bibinfo{author}{\bibfnamefont{U.}~\bibnamefont{St{\"o}sslein}}
  \bibnamefont{and} \bibinfo{author}{\bibfnamefont{E.~R.}
  \bibnamefont{Kinney}}, \bibinfo{journal}{AIP Conf. Proc.}
  \textbf{\bibinfo{volume}{588}}, \bibinfo{pages}{171} (\bibinfo{year}{2001}).

\bibitem[{\citenamefont{Deshpande et~al.}(2005)\citenamefont{Deshpande, Milner,
  Venugopalan, and Vogelsang}}]{Deshpande:2005wd}
\bibinfo{author}{\bibfnamefont{A.}~\bibnamefont{Deshpande}},
  \bibinfo{author}{\bibfnamefont{R.}~\bibnamefont{Milner}},
  \bibinfo{author}{\bibfnamefont{R.}~\bibnamefont{Venugopalan}},
  \bibnamefont{and}
  \bibinfo{author}{\bibfnamefont{W.}~\bibnamefont{Vogelsang}},
  \bibinfo{journal}{Ann. Rev. Nucl. Part. Sci.} \textbf{\bibinfo{volume}{55}},
  \bibinfo{pages}{165} (\bibinfo{year}{2005}).

\end{thebibliography}

\end{document}